# The Influence of Streamlined Music on Cognition and Mood

Julia A. Mossbridge, M.A., Ph.D., jmossbridge@gmail.com
Staff Scientist and Director of the Innovation Lab, Institute of Noetic Sciences
Visiting Scholar, Northwestern University Department of Psychology
Science Director, Focus@Will Labs

*I think I should have no other mortal wants, if I could always have plenty of music. It seems to infuse strength into my limbs, and ideas into my brain. Life seems to go on without effort, when I am filled with music.*
— Mary Ann Evans (George Eliot), *The Mill on the Floss*

*Abstract*[1] –Recent advances in sound engineering have led to the development of so-called "streamlined music" designed to reduce exogenous attention and improve endogenous attention. Although anecdotal reports suggest that streamlined music does indeed improve focus on daily work tasks and may improve mood, the specific influences of streamlined music on cognition and mood have yet to be examined. In this paper, we report the results of a series of online experiments that examined the impact of one form of streamlined music on cognition and mood. The tested form of streamlined music, which was tested primarily by listeners who felt they benefited from this type of music, significantly outperformed plain music on measures of perceived focus, task persistence, precognition, and creative thinking, with borderline effects on mood. In contrast, this same form of streamlined music did not significantly influence measures assessing visual attention, verbal memory, logical thinking, self-efficacy, perceived stress, or self-transcendence. We also found that improvements in perceived focus over a 2+month period were correlated with improvements in emotional state, including mood. Overall the results suggest that at least for individuals who enjoy using streamlined music as a focus tool, streamlined music can have a beneficial impact on cognition without any obvious costs, while at the same time it may potentially boost mood.

*Key Words* – music, streamlined music, cognition, precognition, creative thinking, cognitive flow.

---

[1] See Conflict of Interest and data availability statements at the end of this article.

## BACKGROUND AND MOTIVATION

Music has been used to influence our thoughts and feelings for millennia. In modern times, as we watch a movie, are led to think a phrase just spoken is important because of a shift in the musical score, just as we are led to believe that the heroine is about to meet her enemy at the moment the music changes. Aside from the world of film, in our everyday lives we listen to music at least partially because we believe it will help us concentrate, get things done, and shift our moods (for review, see [1]).

Scientific examinations of the influence of music on cognition and mood are becoming more commonplace, though they still represent a minority of studies examining the influence of sensory input on cognition. Two hypotheses dominate these studies: 1) music will be detrimental to cognitive tasks to the extent that the listener devotes attentional resources to processing the music [2], and 2) music will improve performance on cognitive tasks to the extent that the listener's state of arousal and mood are improved by the music [3]. Which is correct? They are not mutually exclusive hypotheses, and it is likely they are both correct – music helps boost mood and arousal, but if the listener devotes attention to the music, any gain in cognition induced by increased mood and arousal is lost.

Overall, the relationship between music and the mind seems to depend on the task and the music. For example, high-intensity, fast-rhythm versions of a Mozart Sonata hindered reading comprehension [4], while a slower version of the same piece boosted visuo-spatial cognition [3]. Further, familiar music may be more distracting than unfamiliar music [4], although data supporting this idea are sparse. A meta-analysis examining the influence of background music on different kinds of tasks found no effect overall, but did find positive effects on motor-related tasks such as athletic skill [5]. The authors of that meta-analysis asked researchers to examine particular types of music and their effects on particular types of tasks. This paper and the experiment it reports is one response to that request.

We examined a type of music called *streamlined music*, which is electronically recorded music designed with an aim of increasing an individual's focus on cognitive tasks by improving endogenous attention and reducing exogenous attention, essentially supporting the listener in entering a state of flow [6]. We define streamlined music as unfamiliar music to which one or more methods have been applied in order to reduce the activation of exogenous attentional mechanisms in the listener. These methods could include removing frequencies that have been reported to be distracting, or ensuring that the music changes slowly over time. We tested the hypothesis that streamlined music, when played in the background while a listener performs cognitive tasks, produces improvements over plain (i.e., non-streamlined) music.

We predicted that all cognitive measures would be positively influenced by streamlined music, as would mood. We were agnostic about whether streamlined music would influence self-efficacy, perceived stress, or a listener's sense of self-transcendence, but we thought it would be informative to examine self-reports for each of these factors as well, as they are related to the emotional state of a listener. Finally, we hypothesized that improvements in perceived focus ability would support positive shifts in emotional states.

## METHODS

The aim of this Methods section is to give the reader a basic sense of what was done in this experiment; for anyone interested in attempting an exact replication, please contact the author.

*Participants.* We solicited participants through the homepage of a startup that delivers streamlined music for background listening during computer-based work tasks (Focus@Will, http://www.focusatwill.com). Only participants who wanted to participate were given the link to the experiment.[2] Participants were told that we were performing a cognitive research experiment consisting of four testing sessions, and they would receive increasing numbers of free days of the music service in exchange for their participation in each of the tests. Note that this participant payment method eventually screens out participants for whom the service is not desirable (as some of them might not be motivated towards continuing the testing). Thus the conclusions drawn from the experiment described here should only be used to help understand the influence of streamlined music on listeners who enjoy using streamlined music as a focus tool.

*Procedure.* All four testing sessions were conducted online via web software (Word Press and Java), in the homes of the participants, at the time of their choosing. The delay between each of the first three testing sessions was 48 hours to one week. The delay between Testing Session Three and Testing Session Four was 60-80 days. As with any multi-stage study, we found that a decreasing number of participants performed each successive testing session.

*Testing Session One (no background music).* Participants were asked not to listen to any background music during this testing session. We obtained basic demographic information including gender and age. Participants completed a brief Big-5 personality trait inventory [7] and several emotional-state surveys, including the Brief Mood Introspection Survey or BMIS [8], a general self-efficacy questionnaire [9], a perceived stress scale questionnaire [10] and a questionnaire that we had specifically created to assess the experience of self-transcendence or the feeling of connection with something beyond ourselves (available upon request). After completing the emotional-state surveys, participants performed an arrow flanker task [11] and a dimensional change task [12] to acclimate them to cognitive testing. Finally, participants were given a one-item self-report survey to assess their perceived focus on a scale of 1 (low focus) to 5 (high focus).

*Testing Session Two (participant's choice of music or streamlined music).* Half of the participants, randomly selected, were asked to listen to their own choice of background music and the other half were asked to listen to their own choice of streamlined music. Streamlined music was selected from the music channels on the site http://www.focusatwill.com, and plain music could be selected from any other source (e.g., radio, internet music streaming site, digital music files). Participants completed the same mood, general self-efficacy, perceived stress scale, and self-transcendence questionnaires that they completed in Testing Session One. Then they performed the arrow flanker task and the dimensional change task that they performed in Testing Session One, followed by a verbal memory task that was also a test for implicit precognition [13], the Test of Logical Thinking or TOLT, form A [14], and an alternative uses creative thinking task [15]. At the end of this testing session, participants were asked to rank their focus during the session from 1 to 5.

*Testing Session Three (participant's choice of music or streamlined music).* To complete the crossover design, the half of the participants who were asked to listen to their own choice of background music in Testing Session Two were now asked to listen to their own choice of streamlined music, vice versa. The remainder of this testing session was identical with Testing Session Two, except we used TOLT form B [16].

*Testing Session Four (no background music).* Participants were asked not to listen to any background music during this testing session. The purpose of this testing session was to compare scores on self-rated focus and the emotional state surveys to baseline scores on these same measures from Testing Session One.

*Data verification.* Because the experiment was performed online, we could not be sure participants in Testing Sessions One and Four were actually not listening to background music, nor could we be sure participants in Testing Sessions Two and Three were listening to the type of music that was assigned to them. In an attempt to determine whether participants followed instructions, all participants completing both Testing Sessions Two and Three were asked what they listened to during these sessions. Participants who either did not understand the question and therefore could not be expected to understand task instructions, or who responded that they listened to either no music, their choice of music both times, or streamlined music both times were removed from the analysis (10 participants removed in total).

---

[2] Note that while consent forms were not used, this offering the link to the experiment only to participants who pressed a button indicating that they wanted to participate amounts to implied consent.

*Data analysis: Streamlined versus plain music.* To compare the effects of streamlined music to those of plain background music, we performed within-participant comparisons using two-tailed t-tests, except for the data from the combined verbal memory and implicit precognition task (see below). Specifically, we compared data between the two music-listening conditions from participants who completed both Testing Sessions Two and Three.

For the combined verbal memory and implicit precognition task, we only used data from Testing Session Two, which was the first time participants performed this task. This is because this was a task in which words from a word list had to be memorized, and therefore we did not want be concerned about interference from previous word lists, a concern that would have been introduced if we considered data from Testing Session Three. While the combined verbal memory and implicit precognition task was also performed in Testing Session Three, we ignored these data, and simply used the task to keep the tasks and timings between the two testing sessions equivalent.

*Data analysis: Independent validation of perceived focus.* We wanted to compare self-ratings of perceived focus across the two background music conditions (streamlined music versus plain music), but we were aware of a flaw with this plan. Specifically, participants reaching completion of Testing Session Three were likely to be at least partially motivated by receiving an extension of their free trial at the streamlined music service, and therefore would likely be biased toward validating this choice by ranking their experience of focus higher during the streamlined versus plain background music sessions. So we used performance on the TOLT as an objective measure of focus, reasoning that being more focused should produce higher scores on this task. Further, participants did not receive feedback on their performance on the TOLT in either Testing Session Two or Three, and therefore they could not have used this score as a way to assess their experience of focus; thus we could use it as both an objective and independent measure of focus.

*Data analysis: The relationship between changes in perceived focus and emotional state.* To examine whether changes in perceived focus were correlated with changes in positive emotional states, we first calculated difference scores (i.e., Testing Session Four minus Testing Session One) for each of the four emotional state surveys (BMIS, PSS, NGSE, and the Self-Transcendence Scale). We then used linear regression to compare these difference scores to those calculated from the focus survey in the same two testing sessions.

# RESULTS

Nine hundred and nine participants completed the first testing session, 157 of these completed Testing Sessions Two and Three by our deadline, and 50 of those participants completed Testing Session Four by our deadline. Ten of the 157 participants who completed Testing Sessions Two and Three were not able to follow instructions about background music listening and were removed from all analyses. In addition, participant numbers vary slightly across tasks due to some participants not completing each task successfully as a result of technical errors. We note that the bulk of the participants who completed all four testing sessions were those who also were randomly assigned to perform Testing Session Two while listening to streamlined music as opposed to plain music (40 out of 50 participants), a significant effect according to an exact binomial test (probability=0.8, chance=0.5, $p<0.00002$). We can assume these participants were largely motivated to continue testing as a result of their desire to obtain free access to the streamlined music service, but it is not clear why this should be more likely to be the case for participants who heard streamlined music first. We speculate on the interpretation of this unexpected result in *Conclusions*.

**Visual Attention**

We used 30 trials of an arrow-flanker task to assess visual attention. In this task, participants report as quickly as possible the direction of a central arrow (left versus right) that is surrounded by distracting arrows. These arrows can either be pointing in the same direction as the central arrow (congruent) or pointing in the opposite direction (incongruent). The dependent variables are the mean response time (for correct trials) and mean proportion correct on congruent versus incongruent trials.

For the 150 participants who successfully completed the arrow flanker task in both Testing Sessions Two and Three, we found no differences between streamlined music and plain music in either the mean response times or mean proportion correct on congruent and incongruent trials. As expected from the original work on a related flanker task [11], mean response times were slower on incongruent than on congruent trials, but this was the case regardless of background music (see Table 1; streamlined music, $t_{[149]}=7.98$; $p<0.000001$; plain music, $t_{[149]}=8.35$; $p<0.000001$). There was no significant difference between any of the four dependent variables in the two conditions (all $p$-values > 0.057). The only borderline effect was the incongruent condition, on which response times were marginally faster by a mean difference of 20 milliseconds when the background music was plain music as compared to streamlined music.

|  | RT Cong. | RT Inc. | PC Cong. | PC Inc. |
|---|---|---|---|---|
| Streamlined Music | 879 (190) | 943 (178) | 0.993 (0.02) | 0.988 (0.03) |
| Plain Music | 860 (172) | 923 (162) | 0.994 (0.02) | 0.992 (0.03) |

**Table 1.** Means and standard deviations (in parentheses) of response times (RT, in ms) on correct trials and proportion correct values (PC) for congruent (Cong.) and incongruent (Inc.) trials in the arrow-flanker task for assessing visual attention, for both listening conditions. See text for details.

**Task Persistence**

To assess task persistence, we used a dimensional change task consisting of 50 trials. On each trial, just prior to presenting a cartoon image of an animal, we displayed the dimension ("shape" or "color") by which the image was required to be sorted. This task can be used to assess executive function, and specifically cognitive flexibility [12]. This is because after a trial on which participants have sorted the image by one dimension (e.g., "shape"), there is a tendency to continue sorting according to that dimension even if the dimension is changed on the next trial (e.g., "color"). Participants tend to respond more slowly on these "dimension-shift" trials than on trials with the same dimension as the previous trial. The ability to respond accurately and quickly on dimension-shift trials can be seen as a measure of high cognitive flexibility.

Alternatively, because the dimension cue is an exogenous one (a word flashed on the screen prior to each image), and because an endogenous attentional set towards continuing the same task is a type of task persistence, one can view less accurate performance on dimension-shift trials as compared to other trials as a measure of task persistence. On this view, the participant is ignoring or not fully processing the exogenous cue in favor of continuing the previous task. This interpretation would be contingent on visual attention not being negatively affected by the manipulation being examined (which is the case here, see above). However, interpreting increased response times on dimension-shift trials in this same light is not as straightforward as interpreting reduced accuracy, because response times are only calculated for correct trials, and therefore on any trials for which response time is calculated the participants have already made the switch correctly to the new task. Thus here we will interpret lesser accuracy on dimension-shift trials as compared to non-shift trials as an indication of high task persistence.

While listening to both kinds of music, participants were significantly less accurate and slower when responding to dimension-shift trials as compared to other trials (proportion correct for dimension-shift versus non-shift trials: streamlined music, $t_{[148]}=8.06$, $p<0.000001$; plain music, $t_{[148]}=6.03$, $p<0.000001$; response times for dimension-shift versus non-shift trials: streamlined music, $t_{[148]}=5.68$, $p<0.000001$; plain music, $t_{[148]}=6.98$, $p<0.000001$; see Table 2). However, we found the proportion-correct difference score (i.e., proportion correct on dimension-shift trials minus non-shift trials) was significantly higher for streamlined as compared to plain music ($t_{[148]}=2.03$, $p<0.05$, see Figure 1), while there was no significant difference between the response-time difference scores ($p>0.278$). These results, when taken together with the novel interpretation of reduced accuracy on dimension-shift trials, indicate a statistically significant effect in which streamlined music can be thought of as boosting task persistence, potentially via reducing exogenous attention.

|  | RT NonS. | RT DimS. | PC NonS. | PC DimS. |
|---|---|---|---|---|
| Streamlined Music | 1005 (234) | 1058 (234) | 0.985 (0.03) | 0.950 (0.06) |
| Plain Music | 970 (204) | 1036 (208) | 0.986 (0.03) | 0.962 (0.05) |

**Table 2.** Means and standard deviations (in parentheses) of response times (RT, in ms) on correct trials and proportion correct values (PC) for "non-shift" trials on which the dimension was not different from the previous trial (NonS) and "dimension-shift" trials on which the dimension was different from the previous trial (DimS) in the dimensional change task, for both listening conditions. See text for details.

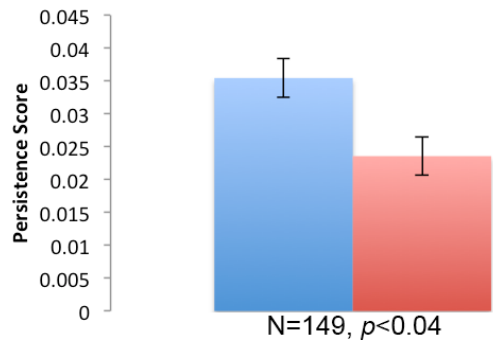

**Figure 1.** Mean task persistence scores in the dimensional change task, as measured by the difference between proportion correct on trials on which the dimension was not different from the previous trial and "dimension-shift" trials on which the dimension was different from the previous trial. Left blue bar: streamlined music; right red bar: plain music. Error bars indicate +/- one standard error of the mean (S.E.M.) within participants.

**Verbal Memory**

In the verbal memory task, participants were shown a list of 48 nouns that were randomly selected for each participant from a set of 96 nouns. These were displayed one at a time in quick succession, and participants were asked to memorize the words in this original word list as well as they could. Then we asked participants to perform a two-alternative forced-choice word-recognition test, in which their goal on each trial was to use their mouse to select which of two words was the one that was in the original word list. We asked them to perform as quickly and accurately as possible. For reasons discussed in *Methods*, we only analyzed data from this task obtained from Testing Session Two (N=144 for streamlined music and N=160 for plain music, all from Testing Session Two; these numbers are larger than for the other tasks because we were not constrained to participants who also performed Testing Session Three). As a result, these comparisons were necessarily between participant groups, so the statistical power of these comparisons is lower than in the within-group analyses.

We found no significant difference between streamlined music and plain music in accuracy or response times on the word-recognition test (both *p*-values>0.206; see Table 3). These results suggest that streamlined music and plain music did not differentially affect verbal memory as assessed in this task.

|  | RT | PC |
|---|---|---|
| Streamlined Music | 1547 (282) | 0.797 (0.18) |
| Plain Music | 1584 (270) | 0.768 (0.21) |

**Table 3.** Means and standard deviations (in parentheses) of response times (RT, in ms) on correct trials and proportion correct values (PC) in the verbal memory task, for participants in either listening condition. See text for details.

**Implicit Precognition**

In addition to being useful for examining verbal memory, we also used the verbal memory task as a test for *implicit precognition*. Implicit precognition is the nonconscious ability to predict future events that should not be predictable [13]. After the participant completed an initial word-recognition test (see *Verbal Memory* above), 24 words from the original word list were randomly selected to be reinforced using training, and these words represented the "future events" that should not have been predictable on the initial word-recognition test, as they had not been selected at the time this test was taken. These words were selected independently for each participant. The dependent variable was the number of these to-be-trained words that participants correctly remembered on the initial word test, as compared to the randomly selected 24 words on the original list that were not to be trained.

Participants practiced the words selected for training in two ways. First, from a list of these 24 words they took four trials to select the words belonging to each of the four categories the nouns represented (people, animals, food, and clothing), a task that required them to attend to the trained words. Second, they saw a picture of each of these 24 words and they were asked to type the appropriate word under each picture. They were not allowed to continue the task until they typed the noun correctly, again requiring them to attend to each of the trained words. Following this training, they were administered a second two-alternative forced-choice word-recognition test on all 48 words in the original word list. The purpose of this second word-recognition test was to determine whether the training had been effective. Thus, participants who recalled fewer trained words than untrained words on this second test were dismissed from the analysis, as we assumed they did not pay attention during the practice portion of the task, or that the training did not work for them. This data-selection step was pre-planned, as the motivation for this step drove the decision to include the second word-recognition test in the first place. We also introduced a non-planned data-selection step, in which we ignored data from participants who scored 90% correct or above on the first word-recognition task, as we reasoned that the memories of these individuals were likely too good to be improved by subtle cues from future training, should such cues exist. After the data from both the streamlined and plain music groups were processed via these two steps, 97 participants remained in each group.

Based on the remaining data, we found that participants listening to streamlined music recognized more to-be-trained words than not-to-be-trained words on the initial word-recognition test ($t_{[96]}$=2.41, $p$<0.018), while this was not the case for participants listening to plain music (see Table 4 and Figure 2). The interaction effect was borderline significant ($t_{[192]}$=1.82, $p$<0.07). Note that prior to the second (not pre-planned) data-selection step, the trend was in the same direction as after the second data-selection step (to-be-trained recall vs. not-to be trained recall, streamlined: $t_{[135]}$=1.63, $p$<0.106; plain music: $t_{[142]}$=-0.51, $p$>0.609)[3].

|  | To-be-trained words | Not-to-be-trained words |
|---|---|---|
| Streamlined Music | 18.52 (4) | 17.90 (4) |
| Plain Music | 18.01 (4) | 18.11 (4) |

**Table 4.** Means and standard deviations (in parentheses) of the number of words out of 24 recognized on the initial word-recognition test in the verbal memory task for the groups of participants remaining after two data-selection steps, for participants in either listening condition. Columns indicate recognition of words to be trained in the future (left column) versus those not to be trained in the future (right column). See text for details.

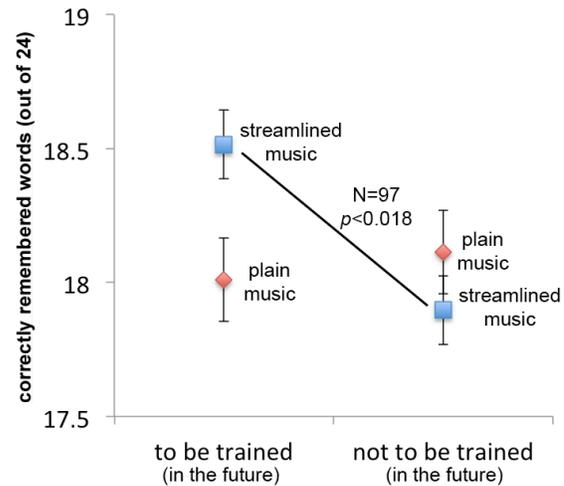

**Figure 2.** Mean numbers of correctly remembered words in the implicit precognition task, for the 24 randomly determined words to be trained in the future (left column) versus the 24 remaining words that would not be trained (right column). Blue symbols give means for participants listening to streamlined music in Testing Session Two, red symbols give means for participants listening to plain music in Testing Session Two. Error bars indicate +/- one S.E.M. within participants.

**Logical Thinking**

To examine logical thinking under both listening conditions, we used the Test of Logical Thinking (TOLT; [14,16]), which assesses mathematical and deductive reasoning. We found no significant difference between listening conditions for scores of the 148 participants who successfully completed this task in both Testing Sessions Two and Three ($t_{[147]}=0.450$, $p>0.653$; see Table 5). This result suggests that background music had no differential effect on logical thinking that could be detected with the TOLT.

---
[3] These data are also presented as part of a review of implicit precognition in a conference proceedings journal that is now in press [13].

|  | TOLT score |
|---|---|
| Streamlined Music | 12.72 (4.14) |
| Plain Music | 12.84 (4.50) |

**Table 5.** Means and standard deviations (in parentheses) of the number of correct items out of 18 on the TOLT for both listening conditions. See text for details.

**Creative Thinking**

To assess creative thinking, we used the alternative uses task [15], in which participants are asked to list as many creative uses of mundane household objects as possible. We used two independent creativity judges to judge the quality and quantity of these uses (i.e., one judge examined only quality without regard to quantity, and the other examined only quantity without regard to quality). These judges were blind to the listening conditions from which their data were obtained. For Testing Session Two, the mundane object was a two-liter bottle and for Testing Session Three, the object was a cardboard paper-towel tube. We found that the two-liter bottle used as the mundane object in Testing Session Two produced more uses, on average, than the cardboard paper-towel tube (regardless of listening condition), introducing concerns about counterbalancing. There were 86 people who performed the while listening to streamlined music in Testing Session Two and 71 who performed this task while listening to plain music in Testing Session Two, so streamlined music was over-represented for the more productive creativity task. To make our analysis more conservative in light of this fact, we only analyzed data from the first 71 participants in each listening group in each test session, to give a total of 142 participants. Note that after this data elimination the results were essentially the same as for the original data, which included all participants who completed both testing sessions.

We found that when participants listened to streamlined music as compared to plain music they produced marginally more alternative uses and had a significantly higher quality of creative uses (number: $t_{[141]}=1.78$, $p<0.078$; quality: $t_{[141]}=3.17$, $p<0.002$; see Table 6 and Figure 3). Additionally, there was a significant and positive linear correlation between quality scores on the alternative uses task performed while listening to plain music with the personality trait of openness ($r=0.173$; $p<0.05$), but this correlation was not present for the quality scores on the same task while listening to streamlined music ($r=0.033$, $p>0.705$).

To examine the relationship between the personality trait of openness and background music more carefully, we performed a median split on openness for the 133 individuals for whom we had successfully recorded personality scores. Examining the creative quality scores of these participants indicated that streamlined music had a significant effect on participants with low openness, but the effect was only borderline significant for participants with high openness (streamlined vs. plain, low openness: $t_{[61]}=3.07$, $p<0.004$; high openness: $t_{[70]}=1.90$, $p<0.07$). Streamlined music boosted the mean quality of creative thinking of participants with low openness above the level of individuals with high openness listening to plain music (Figure 4).

|  | Number | Quality |
|---|---|---|
| Streamlined Music | 7.77 (4.85) | 3.47 (2.05) |
| Plain Music | 7.16 (3.86) | 2.97 (1.58) |

**Table 6.** Means and standard deviations (in parentheses) of the number and quality of creative uses in the alternative uses task, for both listening conditions. Scale for quality was 0-10; number and quality of uses were ranked by separate independent judges blind to the listening condition. See text for details.

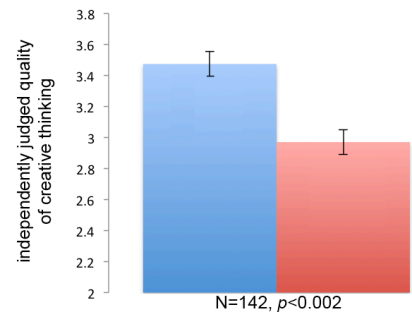

**Figure 3.** Mean quality of creative thinking demonstrated on the alternative uses tasks, measured by an independent judge blind to the listening condition. Blue left column gives the mean for participants as they listened to streamlined music, red right column gives the mean for the same participants as they listened to plain music. Error bars indicate +/- one S.E.M. within participants.

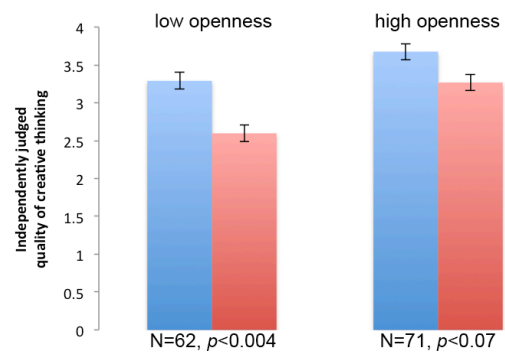

**Figure 4.** Mean quality of creative thinking demonstrated on the alternative uses tasks, measured by an independent judge blind to the listening condition, separated between individuals with openness scores less than 4 on a scale of 1 to 5 (left bars) and those with openness scores of 4 or greater on a scale of 1 to 5 (right bars) based on a median split on the personality trait of openness. Blue columns on the left give the means for participants as they listened to streamlined music, red columns on the right give the means for the same participants as they listened to plain music. Error bars indicate +/- one S.E.M. within participants.

## Emotional State

We obtained complete emotional state assessments from 157 participants who completed both Testing Sessions Two and Three. The average responses on each of the four emotional state surveys were more positive when participants were listening to streamlined music versus when they were listening to plain music (Table 7), though none of the emotional state measures showed any significant difference between listening conditions (all p-values > 0.196). The scale showing the greatest influence from streamlined music was the Brief Mood Introspection Scale (BMIS) for pleasant versus unpleasant mood (Figure 5).

|  | BMIS | PSS* | NGSE | S-T |
|---|---|---|---|---|
| Streamlined Music | 46.20 (7.26) | 2.76 (0.65) | 33.46 (5.49) | 32.47 (5.90) |
| Plain Music | 45.47 (7.28) | 2.77 (0.58) | 33.38 (5.97) | 32.30 (5.73) |

**Table 7.** Means and standard deviations (in parentheses) of responses to the emotional state surveys. BMIS = Brief Mood Introspection Scale, (pleasant vs. unpleasant); PSS = Perceived Stress Scale; NGSE = New General Self-Efficacy Scale; S-T= Self-Transcendence Scale. See text for details. *For PSS, higher scores are negative (greater perceived stress); for all other scales, higher scores are positive.

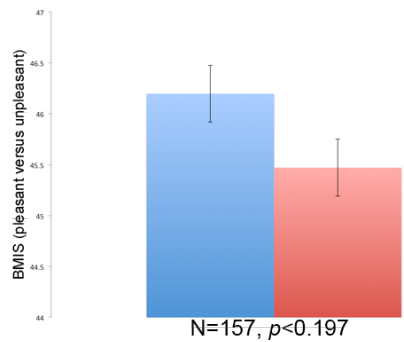

**Figure 5.** Mean of responses on the Brief Mood Introspection Scale (BMIS) for participants as they listened to streamlined music (left blue column) or plain music (right red column). Higher values indicate more pleasant mood. Error bars indicate +/- one S.E.M. within participants.

## Perceived Focus

At the end of each testing session, participants reported their perceived level of focus during the session. We recorded 148 of these self-reported focus responses from participants who performed both Testing Sessions Two and Three. The responses supported the idea that participants felt they were more focused when they were listening to streamlined as opposed to plain music ($t_{[147]}$=4.55, $p$<0.00002; Table 8). As noted in Methods, our process of obtaining participants was likely to bias our sample towards those who felt more focused while listening to streamlined music.

To handle these data more conservatively, we examined whether self-reported focus measures correlated with an objective and independent measure. Because there were no differences between listening conditions on the TOLT (see above), and because this test was the second-to-last test in each testing session (the last test being the alternative-uses task) so that it was relatively close in time to the focus survey, we chose to use scores on the TOLT as an objective measure with which to examine whether participants were accurate in their self-assessment of focus. Note that participants did not receive feedback on their scores for the TOLT, so any correlation between the TOLT scores and perceived focus could only be due to focus being correlated to the TOLT performance itself, not due to participants' knowledge of their performance on the TOLT. We found that among all participants who completed Testing Sessions Two and Three, there was a significant positive correlation between self-rated focus and scores on the TOLT form used in Testing Session Three ($r$=0.313; $p$<0.0002) but the same was not true for scores obtained in Testing Session Two ($r$=0.036, $p$>0.667). Thus if we examine only self-rated focus for Testing Session Three, this more conservative method again reveals that participants listening to streamlined music perceived themselves as significantly more focused than participants who listened to plain music in Testing Session Three ($t_{[146]}$=2.47, $p$<0.015; Figure 6). Note that despite the correlation between perceived focus and the TOLT in Testing Session Three, there was no significant difference between TOLT scores in listening conditions in this testing session.

|  | Focus |
|---|---|
| Streamlined Music | 3.78 (0.80) |
| Plain Music | 3.43 (0.79) |

**Table 8.** Means and standard deviations (in parentheses) of responses to the focus survey in both Testing Sessions Two and Three; higher values indicate greater perceived focus. See text for details.

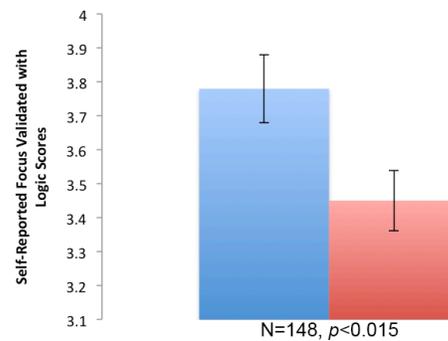

**Figure 6.** Mean of self-reported focus scores in Testing Session 3 on a scale of 1 to 5 for participants; these scores were significantly correlated with scores on the logic test (TOLT). Higher values indicate greater perceived focus. Left blue column shows the mean for participants listening to streamlined music and right red column shows the mean for participants listening to plain music. Error bars indicate +/- one S.E.M. between participants.

## Changes in Self-Reported Focus and Emotional State

We hypothesized that improved focus supports positive changes in emotional state, and vice versa. To partially test this hypothesis, we calculated the changes in self-reported focus scores and emotional state measures between Testing Sessions One and Four for the 46 participants who performed Testing Session Four (see Methods). Note that these two testing sessions were separated by 60-100 days. We found that for three of the four emotional state measures (BMIS, PSS, and the Self-Transcendence Scale), improvements in self-reported focus correlated significantly with improvements in emotional state (Table 9). This finding supports our hypothesis that improved focus supports an improved emotional state and vice versa, though of course causality has not been established.

|  | $\Delta$ BMIS | $\Delta$ PSS* | $\Delta$ NGSE | $\Delta$ S-T |
|---|---|---|---|---|
| correlation versus $\Delta$ focus | $r=0.368$ $p<0.013$ | $r=-0.302$ $p<0.05$ | $r=-0.009$ $p>0.952$ | $r=0.420$ $p<0.004$ |

**Table 9.** Correlations and significance levels for change scores derived from emotional state measures and change scores derived from the focus survey (Testing Session Four minus Testing Session One). BMIS = Brief Mood Introspection Scale, (pleasant vs. unpleasant); PSS = Perceived Stress Scale; NGSE = New General Self-Efficacy Scale; S-T = Self-Transcendence Scale. *For PSS, higher scores are negative (greater perceived stress); for all other scales, higher scores are positive. See text for details.

## CONCLUSIONS AND DISCUSSION

The hypothesis that all cognitive tasks and mood would be positively influenced by streamlined music was partially supported by these data. Specifically, participants had scores that were significantly better on four measures while listening to streamlined, as opposed to plain, music: task persistence, implicit precognition, creative thinking, and perceived focus. The most impressive result was the strong effect on creative thinking, and the most controversial result was the effect on implicit precognition. We discuss both of these findings below. There were no differences between listening conditions in scores on visual attention, verbal memory, logical thinking. While emotional states were generally better while listening to streamlined music, with mood showing the largest effect, these differences were not statistically significant.

There were no tasks on which plain music produced significant decrements in performance as compared to streamlined music, unless the dimensional change task is interpreted as a cognitive flexibility measure instead of a task persistence measure. However, in our case, we were testing the overall hypothesis that streamlined music reduces exogenous attention and improves endogenous attention. This hypothesis instructed our interpretation of the dimensional change information as an exogenous cue, much like a smartphone notification that appears while the goal of proceeding with an ongoing task is being attended via endogenous attention. Even if one is determined to interpret the results on the dimensional change task as representing decreased cognitive flexibility in participants as they listen to streamlined music, the extent of the effect is small enough as to be negligible (percent correct on dimension-switch trials, streamlined music: 95%, plain music: 96%).

Interestingly, we found that the bulk of the participants who completed all four testing sessions were those who also were randomly assigned to perform Testing Session Two while listening to streamlined music as opposed to plain music (40 out of 50 participants), a significant effect according to an exact binomial test (probability=0.8, chance=0.5, $p<0.00002$). We can assume these participants were largely motivated to continue testing as a result of their desire to obtain free access to the streamlined music service. This result may suggest that having the opportunity to first perform cognitive tasks using streamlined music rather than plain music may have made it easier for these participants to believe that streamlined music improved their performance and focus. This could be because these participants could have noticed that when they performed the same tasks again but listening to plain music in Testing Session Three, they did not perform as well as in Testing Session Two. Meanwhile, participants who listened to plain during Testing Session Two could easily decide that practice was the reason that they improved on certain tasks in Testing Session Three, and might not have attributed their improvement to the fact that they were listening to streamlined music in Testing Session Three.

The results from the implicit precognition and creativity measures are intriguing enough to warrant further elaboration. As to implicit precognition, this type of experiment has a rich history that is beyond the scope of this paper (for review, see [17]). However, for the purposes of understanding the present experiment, it is important to note that a previous version of the implicit precognition task used here has proven to be difficult to replicate [18-19]. The authors of a recent meta-analysis propose that one reason for this lack of replication could be that the original version of this task required lengthy deliberation in order to recall the words on the original list, and they had found that implicit precognition tasks requiring much faster responses were more likely to be replicated [19]. As a result of this observation, we adjusted the task so that responses had to be made quickly. But only participants who listened to streamlined music showed the original effect, indicating that perhaps it is not only response speed that influences whether the original effect is replicated or not. Along these lines, it is interesting to note that in that original experiment, participants did the task only after listening to music that had some similarities to streamlined music [18]. Thus it is possible that streamlined music puts people in a state in which subtle cues from future events can be received, processed, and/or used more effectively. Replication is necessary and is underway.

The influence of streamlined music on creative thinking was surprisingly strong. We speculate that the reason for this strong effect is that creative thinking requires the cognitive resources that seem to be differentially recruited by streamlined music. Specifically, streamlined music had a mild positive effect on task persistence; when participants listen to streamlined music they were likely less focused on exogenous cues and more focused on their endogenous attentional set. To maintain the process of searching for creative uses of a mundane task over time undoubtedly requires endogenous attention. In addition, streamlined music improved a measure of implicit precognition, an improvement that can be thought of as the result of better processing of and access to unconscious information. At the same time, creative thinking is supported in situations facilitating unconscious processing and access to unconscious information [20-22], and this may explain the persistent relationship between creativity and openness [23-24]. Taken together, it appears there are some similarities between the mechanisms underlying creativity and implicit precognition. Finally, streamlined music also improved perceived focus, an experience that is often necessary for bursts of creativity, and has been alternately called "flow" [25].

Overall, the results of this study indicate that there is a reasonable empirically based argument for using streamlined music to support focus, creativity, and flow during work. Although this effect may be specific to those who enjoy listening to streamlined music as they work, it is possible that even those who do not enjoy streamlined music could potentially benefit from it. However, the answer to this question is in the domain of future examinations of the influences of streamlined music, and music in general, on cognition and mood.

ACKNOWLEDGMENTS

Daryl Bem provided ideas related to creating the verbal memory/precognition task for online use.

REFERENCES

[1] North, A. C., Hargreaves, D. J., & Hargreaves, J. J. (2004). Uses of music in everyday life. *Music Perception: An Interdisciplinary Journal*, *22*(1), 41-77.

[2] Lavie, N., Hirst, A., De Fockert, J. W., & Viding, E. (2004). Load theory of selective attention and cognitive control. *Journal of Experimental Psychology: General*, *133*(3), 339-354.

[3] Husain, G., Thompson, W. F., & Schellenberg, E. G. (2002). Effects of musical tempo and mode on arousal, mood, and spatial abilities. *Music Perception: An Interdisciplinary Journal*, *20*(2), 151-171.

[4] Thompson, W. F., Schellenberg, E. G., & Letnic, A. K. (2012). Fast and loud background music disrupts reading comprehension. *Psychology of Music*, *40*(6), 700-708.

[5] Kämpfe, J., Sedlmeier, P., & Renkewitz, F. (2010). The impact of background music on adult listeners: A meta-analysis. *Psychology of Music*, 0305735610376261.

[6] Henshall, W. R. (2013). *U.S. Patent Application No. 13/843,585*.

[7] Rammstedt, B., & John, O. P. (2007). Measuring personality in one minute or less: A 10-item short version of the Big Five Inventory in English and German. *Journal of Research in Personality*, *41*(1), 203-212.

[8] Mayer, J. D., & Gaschke, Y. N. (1988). The experience and meta-experience of mood. *Journal of Personality and Social Psychology*, 55, 102-111.

[9] Chen, G., Gully, S. M., & Eden, D. (2001). Validation of a new general self-efficacy scale. *Organizational research methods*, *4*(1), 62-83.

[10] Cohen, S. & Williamson, G. (1988) Perceived Stress in a Probability Sample of the United States. Spacapan, S. and Oskamp, S. (Eds.) *The Social Psychology of Health*. Newbury Park, CA: Sage.

[11] Eriksen, B. A., & Eriksen, C. W. (1974). Effects of noise letters upon the identification of a target letter in a nonsearch task. *Perception & Psychophysics, 16(1), 143-149.*

[12] Zelazo, P. D. (2006). The Dimensional Change Card Sort (DCCS): A method of assessing executive function in children. *Nature Protocols – Electronic Edition, 1*(1), 297.

[13] Mossbridge, J. A. (In press). Examining the nature of retrocausal effects in biology and psychology. *American Institute of Physics Conference Proceedings*.

[14] Tobin, K. G., & Capie, W. (1981). The development and validation of a group test of logical thinking. *Educational and Psychological Measurement*, *41*(2), 413-423.

[15] Torrance, E. P. (1972). Predictive validity of the Torrance Tests of Creative Thinking. *The Journal of Creative Behavior*. Vol 6(4), 236-252

[16] Mann, A. E. (1982). The effects of a problem-solving strategy on the long-term memory of algorithms. *UNF Theses and Dissertations*. Paper 16.

[17] Mossbridge, J. A. & Baruss, I. (2016). *Transcendent mind: Rethinking the science of consciousness*. Washington, D.C.: American Psychological Association Books.

[18] Bem, D. J. (2011). Feeling the future: experimental evidence for anomalous retroactive influences on cognition and affect. *Journal of Personality and Social Psychology, 100*(3), 407.

[19] Bem, D., Tressoldi, P., Rabeyron, T., & Duggan, M. (2015). Feeling the future: A meta-analysis of 90 experiments on the anomalous anticipation of random future events. *F1000Research, 4*.

[20] Zhong, C. B., Dijksterhuis, A., & Galinsky, A. D. (2008). The merits of unconscious thought in creativity. *Psychological Science*, *19*(9), 912-918.

[21] Bowden, E. M., & Beeman, M. J. (1998). Getting the right idea: Semantic activation in the right hemisphere may help solve insight problems. *Psychological Science*, *9*(6), 435-440.

[22] Baird, B., Smallwood, J., Mrazek, M. D., Kam, J. W., Franklin, M. S., & Schooler, J. W. (2012). Inspired by distraction mind wandering facilitates creative incubation. *Psychological Science*.

[23] Feist, G. J. (1998). A meta-analysis of personality in scientific and artistic creativity. *Personality and Social Psychology Review*, *2*(4), 290-309.

[24] Feist, G. J., & Barron, F. X. (2003). Predicting creativity from early to late adulthood: Intellect, potential, and personality. *Journal of Research in Personality*, *37*(2), 62-88.

[25] Csikszentmihalyi, M. (1996). Flow and the psychology of discovery and invention. *New York: Harper Collins*.